\documentclass[twocolumn,showpacs,aps,floatfix,prd,preprintnumbers]{revtex4}
\usepackage{amssymb}
\usepackage{amsmath}
\usepackage{eurosym}
\usepackage{epstopdf}
\usepackage{capt-of}
\usepackage{graphicx}
\usepackage{dcolumn}
\usepackage{hyperref}
\usepackage{bm}
\usepackage{float}

\begin{document}

\title{Viability of the $R+e^T$ cosmology}
\author{ P.H.R.S. Moraes$^{\dagger}$, P.K. Sahoo$^{\ddagger}$, S.K.J. Pacif$%
^{*}$}
\affiliation{$^{\dagger}$Universidade de S\~ao Paulo, Instituto de Astronomia, Geof\'isica e Ci\^encias Atmosf\'ericas, R. do Mat\~ao 1226, Cidade Universit\'aria, 05508-090 S\~ao Paulo, SP, Brazil \footnote{%
Email: moraes.phrs@gmail.com}}
\affiliation{$^{\ddagger}$Department of Mathematics, Birla Institute of Technology and
Science-Pilani, \\
Hyderabad Campus, Hyderabad-500078, India \footnote{%
Email: pksahoo@hyderabad.bits-pilani.ac.in}}
\affiliation{$^*$Department of Mathematics, School of Advanved Sciences, VIT University,
Vellore 632014, Tamil Nadu, India \footnote{%
Email: shibesh.math@gmail.com}}

\begin{abstract}
We propose a new theoretical approach for a cosmological model, which starts
from an exponential of the trace of the energy-momentum tensor-dependence on
the gravitational action, to be summed to the Ricci scalar. We derive the
referred field equations and Friedmann-like equations. We derive the scale
factor, Hubble parameter and deceleration parameter, in terms of both time
and redshift. In possession of those parameters in terms of the redshift, we
confront their predictions with the observational Hubble dataset and the
outcomes are pretty satisfactory so that the model can be seen as a new
alternative to the cosmological constant problem. We also present the
statefinder diagnostic and discuss profoundly the dynamical behaviour of the
model.
\end{abstract}

\maketitle

%\affiliation{$^\dagger$Universit\`a di Napoli ``Federico II'' - Dipartimento di Fisica,
%Napoli I-80126, Italy\\

%\input epsf.tex %%%%%%%%%%%%
%%%%%%%%%%%

%%%%%

%%%%%%%%%%%%%%%%%%%%%%%%%%%%%%%%%%%%%%%%%%

\section{Introduction}\label{sec:int}

The greatest mystery of Physics nowadays is the cause of the acceleration of
the expansion of the universe. The expansion of the universe, discovered by
E. Hubble in 1929 \cite{hubble/1929}, was expected to be slowing down due to
the attractive feature of the force of gravity. However, at the end of the
20th century, the measuring of the brightness of distant supernovae Ia \cite%
{riess/1998,perlmutter/1999} has shown that the expansion is, in fact, speeding
up. Approximately 20 years later, what causes this acceleration still remains
unknown.

While the standard cosmological model, named $\Lambda$CDM model, for which $%
\Lambda$ represents the cosmological constant and CDM stands for \textit{%
cold dark matter}, indeed provides a great fit with observations \cite%
{planck_collaboration/2017}, there is a strong shortcoming in what concerns
the physical interpretation of the cosmological constant \cite{weinberg/1989}%
-\cite{padmanabhan/2003}. Neither have dark matter particles been detected 
\cite{bednyakov/2016,mambrini/2007}.

Since $\Lambda$CDM model is based on Einstein's General Theory of Relativity 
\cite{einstein/1915}, theoretical physicists have been trying to solve the
above issues by extending General Relativity. The idea is that some new
degrees of freedom of extended theories of gravity could play the role of
the dark sector. In fact, some extra terms in the field equations of
extended gravity theories \cite{de_felice/2010,capozziello/2011} indeed can
describe the dark energy and matter effects simply as correction terms to
General Relativity.

The dark energy and matter dynamical effects have been obtained within
extended gravity theories as one can check References \cite%
{nojiri/2006,shirasaki/2016} and \cite{capozziello/2013,rodrigues/2014},
respectively. Naturally, those theories may also have some shortcomings. For
instance, solar system tests have ruled out most of the $f(R)$ models
proposed so far \cite{chiba/2003,nojiri/2008}, for which $f(R)$ means a
general function of the Ricci scalar, to replace $R$ in the usual
Einstein-Hilbert action.

Our proposal herewith in the present paper is to address the cosmic
acceleration in an extended gravitational theory that allows the
generalization of the material sector of the field equations of General
Relativity, rather than their geometrical sector.

We are going to replace $R$ in the Einstein-Hilbert gravitational action by $%
R+f(T)$, with $T$ representing the trace of the
energy-momentum tensor. That is to say that we are going to work with a particular case of the $f(R,T)$ gravity \cite{harko/2011}. The $T$-dependence in a gravitational theory is related to the description of some quantum effects (conformal anomaly) \cite{harko/2011}, as we are going to revisit later on. It is also motivated by the possible existence of imperfect fluids in the universe. In fact, when putting $f(T)$ in the gravitational action, the field equations of the model can be cast in terms of an effective energy-momentum tensor whose extra terms are related to imperfections such as anisotropy, viscosity, elasticity etc.

The function $f(T)$ is, in principle, arbitrary, as the $f(R)$ function in
the $f(R)$ gravity case. We are going to suppose here an exponential
dependence for $T$ as $f(T)\sim e^{\chi T}$, with $\chi $ being a free
parameter. Such an assumption is somehow motivated by exponential $f(R)$
gravity \cite{odintsov/2017}. Anyhow, the assumption of dynamical fields
exponentially entering the gravitational action is not an exclusivity of the
exponential $f(R)$ gravity. In Ref. \cite{pettorino/2005}, a scalar
field is playing the role of dark energy and is coupled to the Ricci scalar
exponentially. In Ref. \cite{harko/2010}, it was proposed the
substitution of $R$ by $\Lambda e^{\frac{R+L}{\Lambda }}$ in the
gravitational action, with $L$ being the matter Lagrangian.

Our article is organized as follows: in Section \ref{sec:etg}, we present
the basic formalism of $R+e^{T}$ gravity and in Section \ref{sec:etc}, we discuss the $R+e^{T}$ cosmology with a simple exact solution. The dynamical behaviour of the obtained model in the presented $R+e^{T}$ cosmology is further discussed in Section \ref{sec:CD}. In Section \ref{sec:cwo}, we confront the
theoretical predictions of our model with observational Hubble dataset. In
Section \ref{sec:sd}, we present the statefinder diagnostic of our model.
Our final remarks and discussion are presented in Section \ref{sec:d}.

\section{$R+e^T$ gravity}\label{sec:etg}

The field equations of $e^T$ gravity will be obtained from the variation of
the action

\begin{equation}  \label{etg1}
S=\int d^4x\sqrt{-g}\left(\frac{R+\gamma e^{\chi T}}{16\pi}+L\right),
\end{equation}
with $g$ being the metric determinant, $\gamma$ a constant and natural units
are assumed. Such a variation with respect to the metric $g_{\mu\nu}$ yields

\begin{equation}  \label{etg2}
G_{\mu\nu}=8\pi T_{\mu\nu}^{\mathtt{eff}},
\end{equation}
with

\begin{equation}  \label{etg3}
T_{\mu\nu}^{\mathtt{eff}}=T_{\mu\nu}+\frac{\gamma e^{\chi T}}{8\pi}\left[%
\frac{g_{\mu\nu}}{2}+\chi(T_{\mu\nu}+pg_{\mu\nu})\right],
\end{equation}
with $T_{\mu\nu}$ being the usual energy-momentum tensor of matter and $p$
the pressure.

The covariant derivative of the usual matter energy-momentum tensor in %
\eqref{etg3} reads

\begin{multline}  \label{etg4}
\nabla^\mu T_{\mu\nu}=\\
-\frac{\gamma\chi e^{\chi T}}{8\pi+\gamma\chi e^{\chi T}}\left[%
\chi(T_{\mu\nu}+pg_{\mu\nu})\nabla^\mu T+\nabla^\mu\left(\frac{T}{2}%
+p\right)g_{\mu\nu}\right].
\end{multline}

From \eqref{etg4}, it can be seen that the energy-momentum tensor is not
conserved in the present formalism. Such a feature can also be seen in other
models \cite{harko/2010}-\cite{mignani/1997}. It has been shown that it can
be related to a process of particles creation \cite{harko/2015,harko/2014}.

Furthermore, from a completely different approach, in the cosmological
models proposed in \cite{steigman/2009}-\cite{lima/2012}, the mechanism
behind the acceleration of the universe expansion is exactly the production
of particles.

\section{$R+e^T$ cosmology}\label{sec:etc}

With the set up of $R+e^{T}$ gravity discussed above, we now describe the $R+
e^{T}$ cosmology for which, we consider the well-known Friedmann-Lema\^{\i}tre-Robertson-Walker metric \cite{friedmann/1922}-\cite{walker/1937}
describing an isotropic and homogeneous universe. We make null the curvature of  the spatial sector of the space-time, in
accordance with observations of fluctuations of temperature in the cosmic
microwave background radiation \cite{planck_collaboration/2017}.

In parallel, we will consider the energy-momentum tensor of a perfect fluid.
Also, in what concerns the material content of the universe in the present
formalism, we desire to check if it is possible to describe the acceleration
of the universe expansion as a consequence of the correction terms of the
present gravitational model rather than as due to the presence of some
exotic fluid such as dark energy permeating the universe. Therefore we
should assume $p=0$ in the dynamical equations. This is equivalent to say
that dust is presently the dominant component of the universe regulating the
dynamics.

The above proceeding yields

\begin{eqnarray}
3H^2=8\pi\rho+\gamma e^{\chi\rho}\left(\frac{1}{2}+\chi\rho\right),
\label{etc1} \\
-2\dot{H}=(8\pi+\gamma\chi e^{\chi\rho})\rho  \label{etc2}
\end{eqnarray}
and

\begin{equation}  \label{etc3}
\dot\rho\left[\frac{\gamma\chi e^{\chi\rho}}{8\pi+\gamma\chi e^{\chi\rho}}%
\left(\frac{1}{2}+\chi\rho\right)+1\right]+3H\rho=0.
\end{equation}
In Eqs.\eqref{etc1}-\eqref{etc3} above, $H=\dot{a}/a$ is the Hubble
parameter, $a$ is the scale factor, which dictates how distances evolve in
the universe, and dots represent time derivatives. Moreover, it is important
to notice that the usual Friedmann equations for pressureless matter (in the
absence of the cosmological constant) are retrieved by taking $\gamma=0$.

From Equations (\ref{etc1}) and (\ref{etc2}) we get 
\begin{equation}  \label{etc4}
\rho=\log \biggl(\frac{2\mathcal{H}}{\gamma}\biggr)^{\frac{1}{\chi}},
\end{equation}
\begin{equation}\label{etc0}
\dot{\rho}=\frac{2(\ddot{H}+3H\dot{H})}{\chi\mathcal{H}},
\end{equation}
in which it was defined $\mathcal{H}\equiv2\dot{H}+3H^{2}$.

Using Equations (\ref{etc4}) and (\ref{etc0}) into Equation (\ref{etc3}) we
obtain a dynamical equation in the form,

\begin{multline}  \label{etc5}
2\gamma(\ddot{H}+3\dot{H}H)\left\{1+\frac{2\chi\mathcal{H}^2\left[\frac{1}{2}%
+\log\left(\frac{2\mathcal{H}}{\gamma}\right)\right]}{8\pi+2\chi\mathcal{H}}%
\right\} \\
+3\log(2\mathcal{H}^2)H=0.
\end{multline}

A critical inspection of Equation (\ref{etc5}) renders a simple solution for
the condition

\begin{equation}
\mathcal{H}>0,  \label{etc6}
\end{equation}%
which yields the scale factor $a(t)$ in the form of a hybrid expansion law
(power and exponential laws) as,

\begin{equation}
a(t)=\beta t^{\frac{2}{3}}e^{\alpha t},
\end{equation}
with $\alpha$ and $\beta$ being integrating constants.

The referred Hubble and deceleration parameters, with the latter defined as $%
q=-\ddot{a}a/\dot{a}^2$, such that negative values indicate an accelerated
expansion, are

\begin{equation}
H(t)=\alpha +\frac{2}{3t},  \label{h}
\end{equation}%
\begin{equation}
q(t)=\frac{-3\alpha t(4+3\alpha t)+2}{(2+3\alpha t)^{2}}.  \label{dp}
\end{equation}

The cosmic dynamics of the obtained model can be discussed now.

\section{Cosmic Dynamics}\label{sec:CD}

With the exact solutions for the field equations in the $R+e^{T}$ cosmology, we shall now discuss the dynamics of the particular model obtained here. A quick
analysis of the behavior of these geometrical parameters near the initial
singularity and in the far future can be described in the following table

\begin{table}[h!]
%\centering
\begin{center}
\begin{tabular}{|c|c|c|c|}
\hline
Parameter & $a$ & $H$ & $q$ \\ \hline
as $t\rightarrow 0$ & $0$ & $\infty $ & $1/2$ \\ \hline
as $t\rightarrow \infty $ & $\infty $ & $\alpha $ & $-1$ \\ \hline
\end{tabular}
\end{center}
\caption{Limiting values of geometrical parameters}
\label{table:1}
\end{table}

From the above table \ref{table:1}, we can observe that in this model the universe starts
with a singularity and rapidly expands but with
decreasing rate of expansion, which is suitable for structure formation. Also, it behaves like a de Sitter universe in the far future. 

We can also plot these cosmological parameters w.r.t. cosmic time $t$ (in units of billions of years).

From the plot of the scale factor in Figure 1, we can see two different types of
evolution with the choice of the model parameters $\alpha$ and $\beta$. The Hubble parameter in Figure 2 is almost the same independently of the $\alpha,\beta$ values and decreases rapidly. The rate of expansion can be seen in the plot of the deceleration parameter in Figure 3 for these choices of model parameters. For, the best fit values, in the following we shall constrain these model parameters with observations. 

\begin{figure}[tbp]
\centering
\includegraphics[scale=0.3]{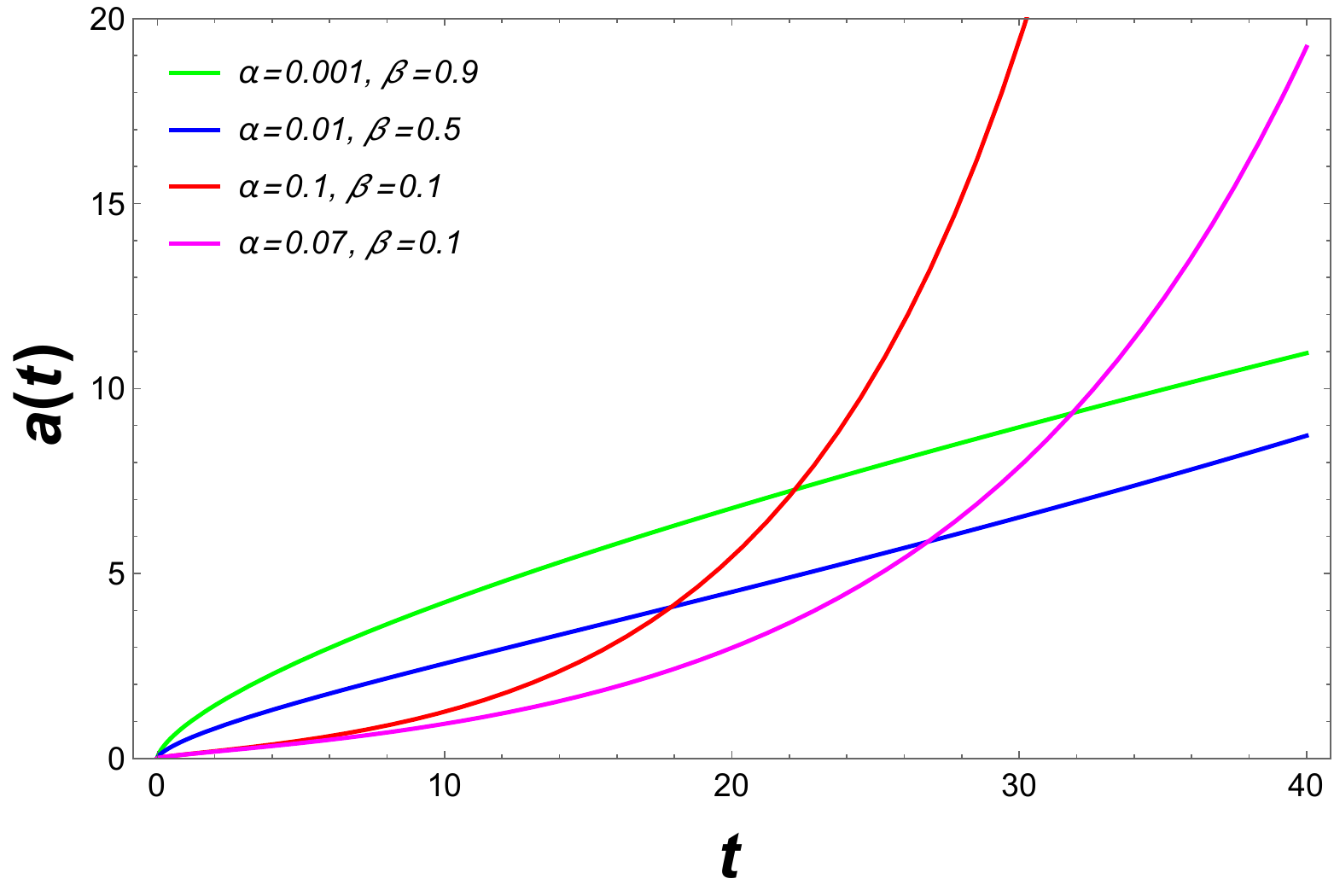}
\caption{{Evolution of the scale factor $a(t)$ w.r.t cosmic time.}}
\label{fig:a-t}
\end{figure}

\begin{figure}[tbp]
\centering
\includegraphics[scale=0.3]{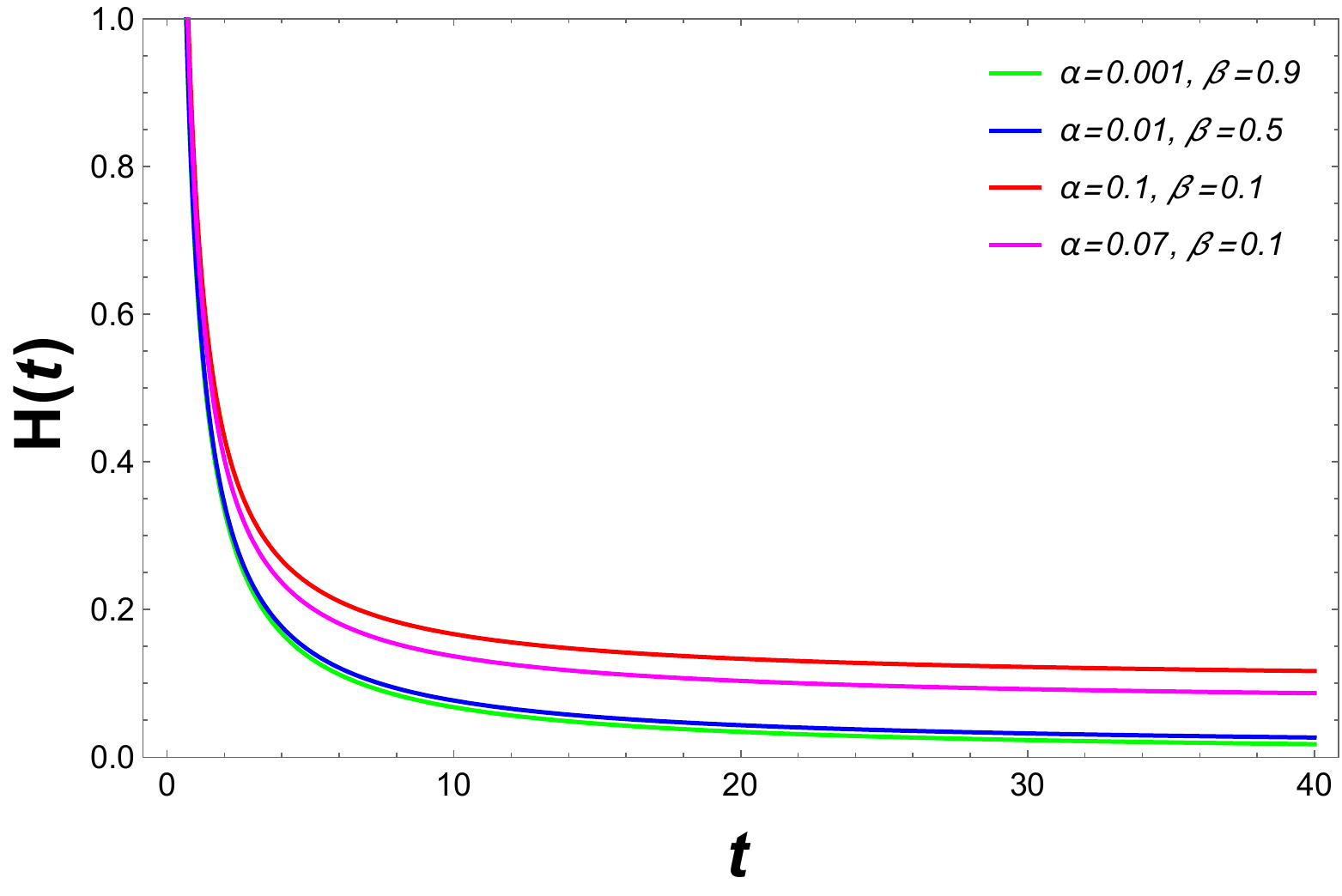}
\caption{{Evolution of the Hubble parameter $H(t)$ w.r.t cosmic time.}}
\label{fig:H-t}
\end{figure}

\begin{figure}[tbp]
\centering
\includegraphics[scale=0.3]{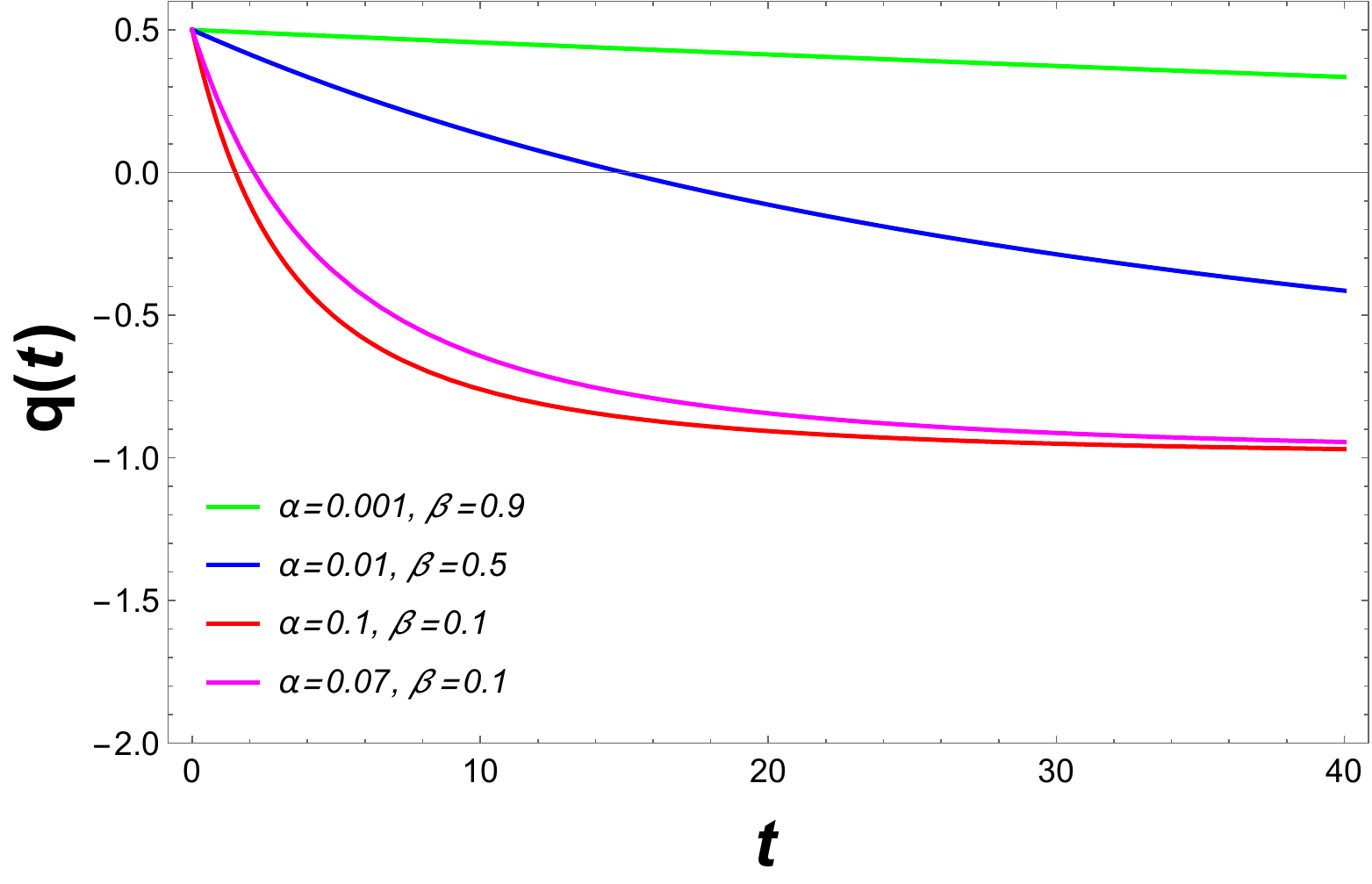}
\caption{{Evolution of the deceleration parameter $q(t)$ w.r.t cosmic time.}}
\label{fig:q-t}
\end{figure}

Further, the density parameter takes the form

\begin{equation}
\rho =\log \left\{\left[\frac{2\alpha (3 \alpha t+4)}{\gamma t}%
\right]^{1/\chi }\right\}  \label{rh}
\end{equation}
(recalling we are assuming that dust matter (with $p=0$) is filling the universe). 

In order to confront our results with observations, it is convenient to rewrite all of these cosmological parameters in terms of the redshift $z$ rather than in terms of time. To do so, we use $a(t)=\frac{a_{0}}{1+z}$, with the present value of the scale factor $a_{0}=1$. The time-redshift relation for our obtained model can be expressed as

\begin{equation}
t=\frac{2}{3\alpha}\mathcal{W}\left\{\frac{3\alpha}{2}\frac{1}{%
[\beta(1+z)]^{3/2}}\right\},
\end{equation}%
where $\mathcal{W}$ denotes the Lambert function, also known as
\textquotedblleft product logarithm\textquotedblright. For the sake of
simplicity, we will take, from now on, $\Gamma\equiv\frac{3\alpha}{2}\frac{1%
}{[\beta(1+z)]^{3/2}}$.

Eqs.\eqref{h} and \eqref{dp} now read

\begin{equation}
H(z)=\alpha\left[1+\frac{1}{\mathcal{W}(\Gamma)}\right],
\end{equation}
\begin{equation}
q(z)=\frac{1}{2}\frac{1-2\mathcal{W}(\Gamma)[2+\mathcal{W}(\Gamma)]}{[1+%
\mathcal{W}(\Gamma)]^2}.
\end{equation}

\section{Confrontation with Hubble data}\label{sec:cwo}

In this section, we put constrains on the model parameters $\alpha $ and $%
\beta $ of the presented model in view of the observational Hubble dataset
(OHD) $H(z)$.

We consider the OHD containing $57$ data points of $H$ in the redshift range 
$0.07\leqslant z\leqslant 2.36$ with the corresponding standard deviations $%
\sigma _{H}$. $31$ points are obtained by the differential age
techniques applied to evolving galaxies and $26$ points are obtained by BAO
and other methods \cite{Hz}. We take the present value of the Hubble
parameter as $H_{0}=67.8$Km/s/Mpc, as suggested by Planck satellite results 
\cite{planck_collaboration/2015} to complete the dataset. In the
calculation of $H(z)$ dataset, the mean values of the model parameters $%
\alpha $ and $\beta $ are determined by minimizing 

\begin{equation}
\chi _{OHD}^{2}(p_{s})=\sum\limits_{i=1}^{57}\frac{%
[H_{th}(p_{s};z_{i})-H_{obs}(z_{i})]^{2}}{\sigma _{H(z_{i})}^{2}},
\end{equation}%
where $p_{s}$ denotes the parameters of the model to be constrained, $H_{th}$
denotes the theoretical (model based) values of the Hubble parameter, $H_{%
\text{obs}}$ signifies the observed values of the Hubble parameter and $%
\sigma _{H(z_{i})}$ is the standard error in each observed value. The
summation runs over $57$ observational data points at redshifts $z_{i}$
together with the $H_{0}$ value.

\begin{figure}[]
\centering
\includegraphics[scale=0.5]{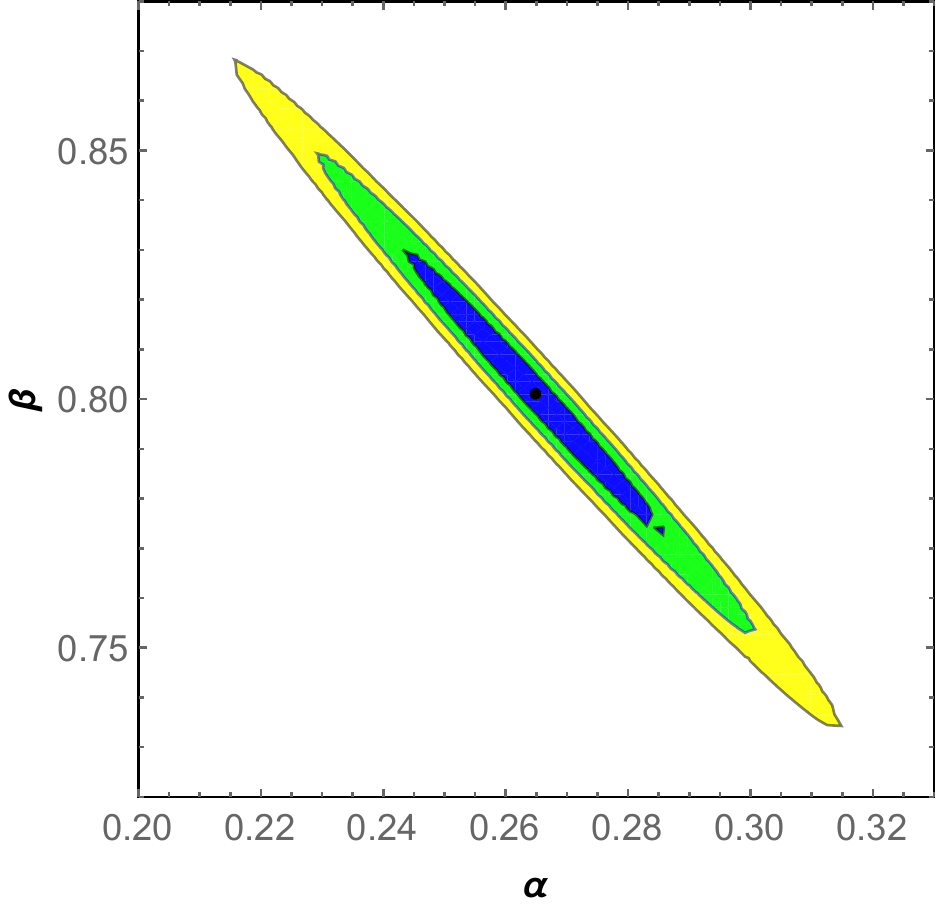}
\caption{{Plots for $1\protect\sigma $ (blue shaded), $2\protect\sigma $
(green shaded) and $3\protect\sigma $ (yellow shaded) likelihood contours in
the $\protect\alpha $-$\protect\beta $ plane for the $R+e^T$ cosmological
model.}}
\label{fig:Contour-EXP}
\end{figure}

The likelihood contours in the $\alpha -\beta $ plane with $1\sigma $, $%
2\sigma $ and $3\sigma $ errors are obtained for our model as shown in Fig.%
\ref{fig:Contour-EXP}. The best fit values of $\alpha $ and $\beta $ come
out to be $\alpha =0.265005$ and $\beta =0.800925$, with minimum $\chi
^{2}=31.6247$. 
%The contour plot with the systematic errors at $1\sigma,2\sigma $ and $3\sigma $ level is shown in Fig.\ref{fig:Contour-EXP}.

\begin{figure}[tbp]
\centering
\includegraphics[scale=0.5]{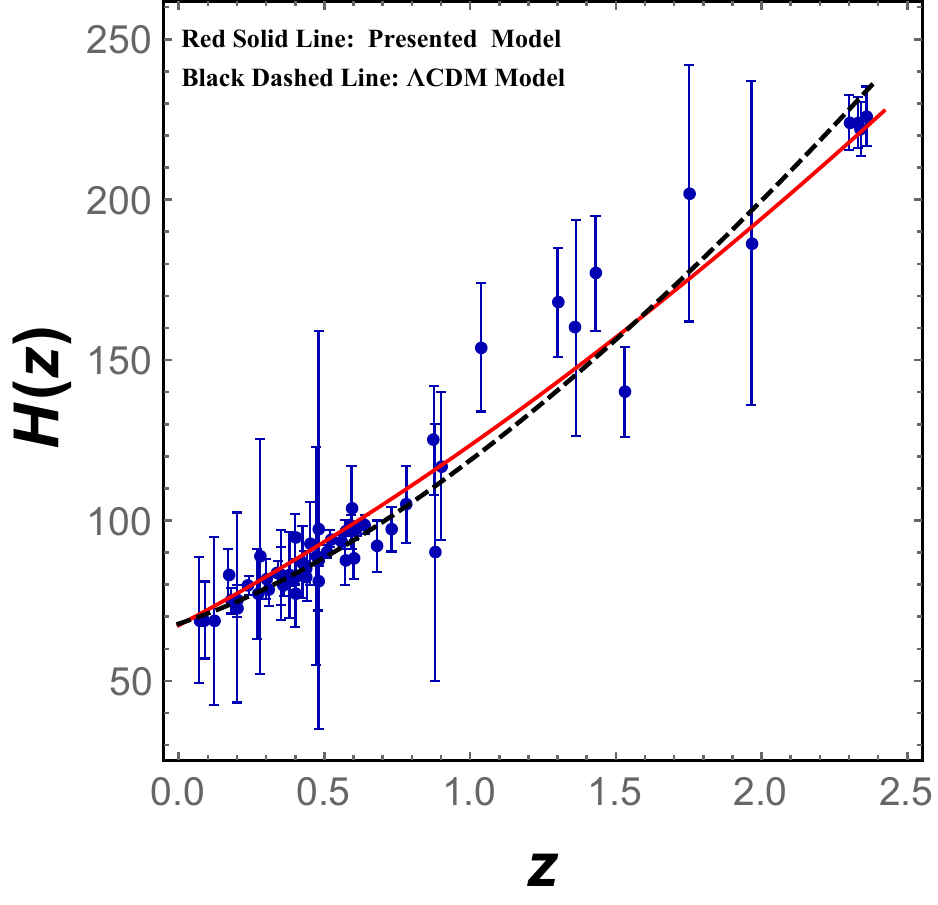}
\caption{${57}${\ points of $H(z)$ dataset with error bars along with the
present model (solid red line) and $\Lambda $CDM model (black dashed line).}}
\label{fig:hzerror}
\end{figure}

The Hubble data with error bars can be seen in Fig.\ref{fig:hzerror}
together with the curves of the $\Lambda $CDM and $R+e^T$ models. The figure
clearly shows a nice fit to the OHD for the $R+e^T$ model with $\alpha=0.265005
$ and $\beta =0.800925$.

\begin{figure}[]
\includegraphics[scale=0.35]{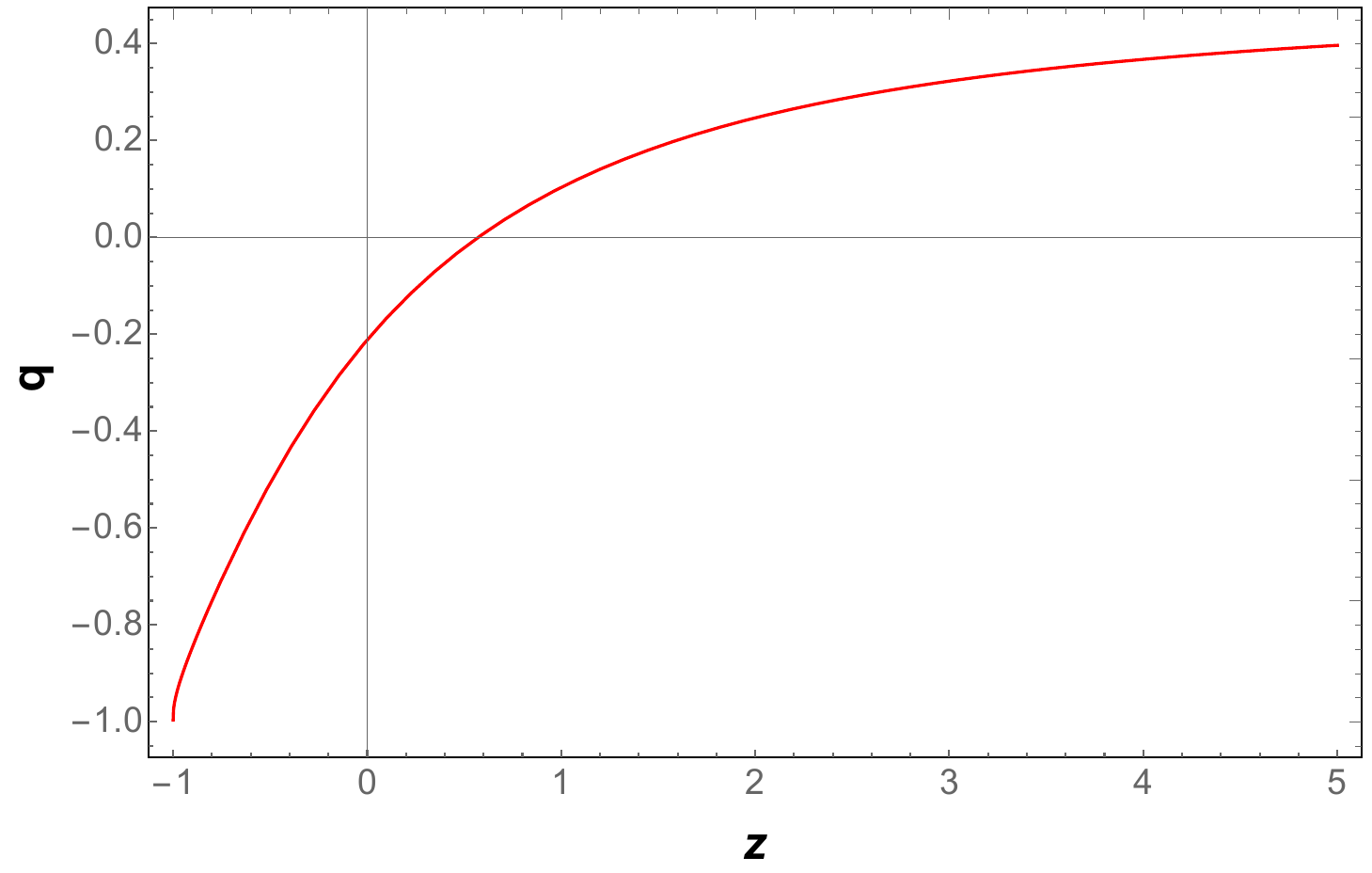}
\caption{Deceleration parameter vs. redshift, with $\protect\alpha =0.265005$
and $\protect\beta =0.800925$.}
\label{fig1}
\end{figure}

One of the most important and crucial phenomena in the evolution of the
Universe is the cosmological \textquotedblleft phase
transition\textquotedblright\ i.e. the time when the universe changes from a
decelerating phase of expansion to an accelerating one. The model we
obtained here indeed possesses the characteristic of cosmological phase
transition for the constrained values of $\alpha $ and $\beta $ as one can see 
in Fig.\ref{fig1}.

From Fig.\ref{fig1}, one can observe that the phase transition from
deceleration to acceleration occurs at $z_{t}=0.571166$. $z_{t}$ for our
model and is in good agreement with the value predicted in \cite%
{Capozziello14}-\cite{Farooq17}. Moreover, we can calculate the present
value of the deceleration parameter as $q_{0}=-0.22$ and is also in
agreement with some recent constraints put on $q_{0}$ \cite{giostri/2012}.

With the constrained values of the model parameters $\alpha$ and $\beta$, the plot of $\rho (z)$ is shown in Figure 7.

\begin{figure}[tbp]
\centering
\includegraphics[scale=0.35]{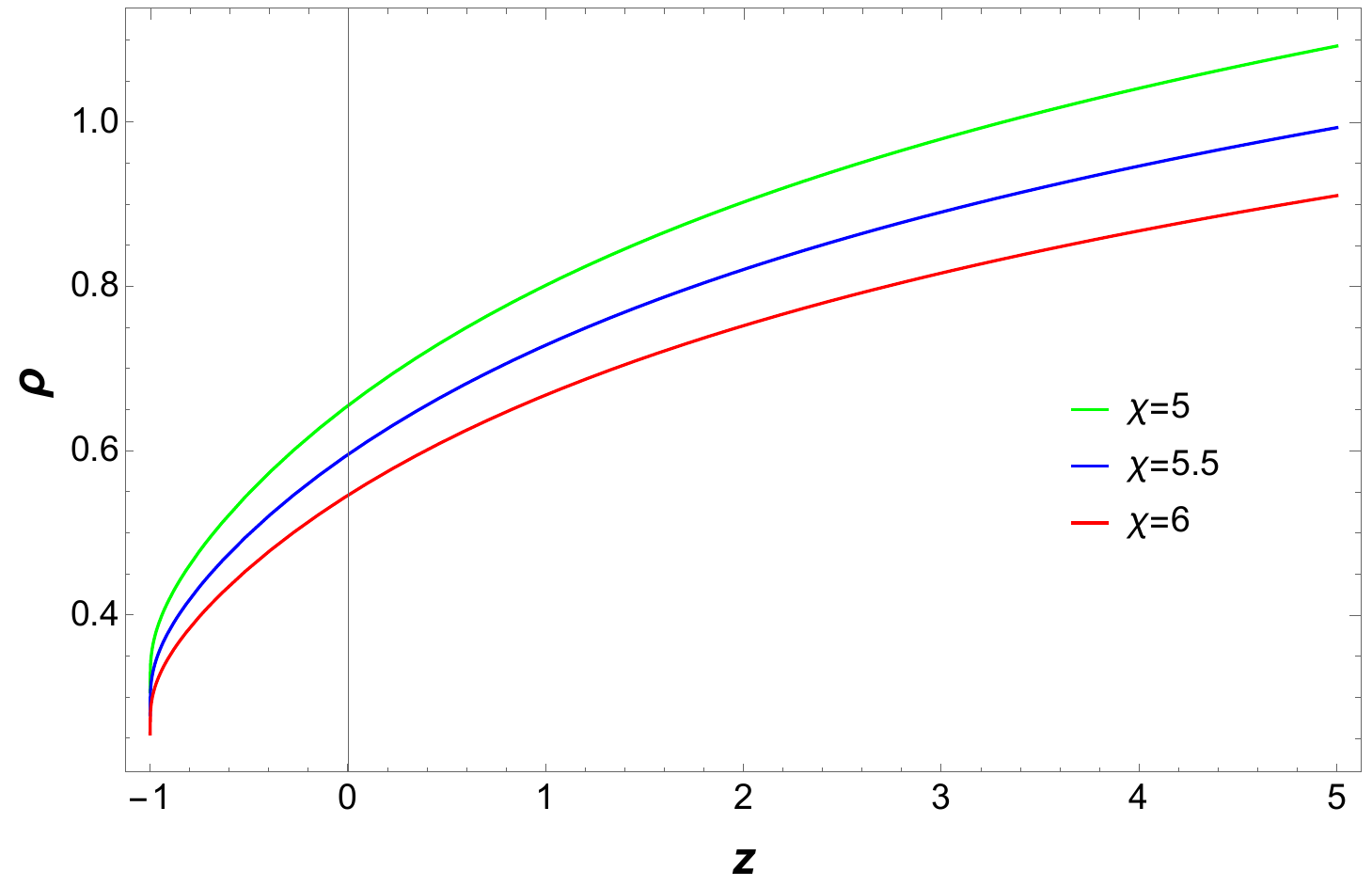}
\caption{Density vs. redshift, with $\protect\alpha =0.265005$ and $\protect%
\beta =0.800925$.}
\label{fig:rho}
\end{figure}

\section{Statefinder diagnostic}\label{sec:sd}

The necessity of considering more general dark energy models than the
standard one together with the increasing in the accuracy of cosmological
observations has led V. Sahni and collaborators to introduce a cosmological
diagnostic pair, namely statefinder pair $\{r,s\}$ \cite{Sahni/2003,
Alam/2003}. This is constructed from the scale factor and its derivatives
up to the third order and reads

\begin{equation}
r=\frac{\dddot{a}}{aH^{3}},
\end{equation}
\begin{equation}
s=\frac{1}{3}\left(\frac{-1+r}{-\frac{1}{2}+q}\right),
\end{equation}
where $q\neq \frac{1}{2}$.

An useful way of distinguishing different cosmological
models that have similar kinematics is to plot the evolution trajectories
of the ${q-r}$ and ${r-s}$ pairs.

For our model, the pair $r,s$ reads

\begin{equation}
r=2\frac{4+9\alpha t[-1+\alpha t(2+\alpha t)]}{(2+3\alpha t)^{3}},
\end{equation}
\begin{equation}
s=\frac{4}{8+9\alpha t(2+\alpha t)}.
\end{equation}

\begin{figure}[]
%\centering
\includegraphics[scale=0.5]{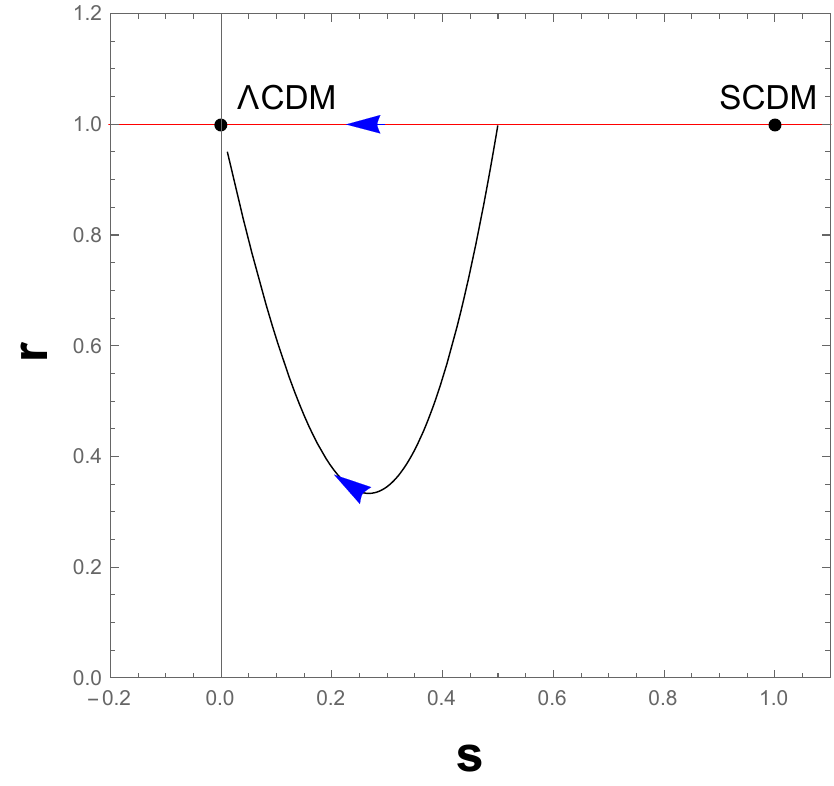}
\caption{Variation of $s$ vs. $r$ for the $R+e^T$ cosmological model with $%
\protect\alpha=0.265005$ and $\protect\beta=0.800925$.}
\label{fig7}
\end{figure}

\begin{figure}[]
%\centering
\includegraphics[scale=0.5]{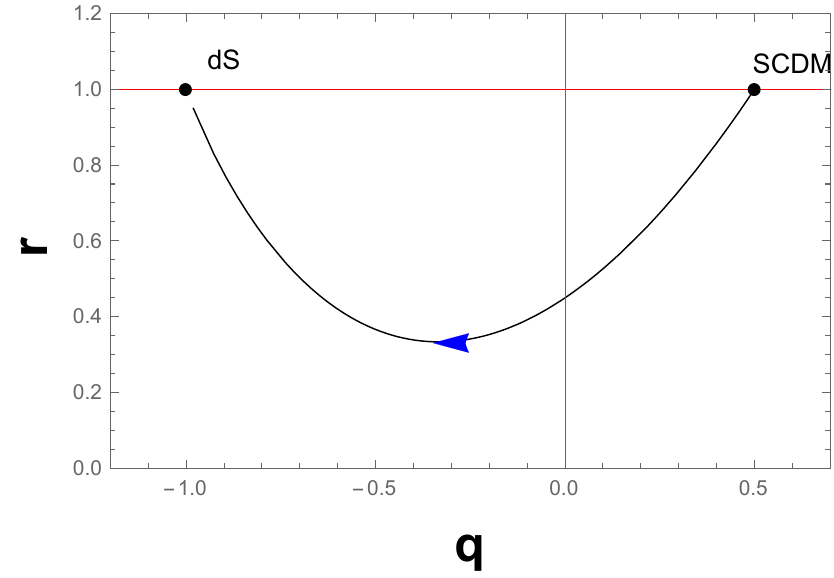}
\caption{Variation of $q$ vs. $r$ for the $R+e^T$ cosmological model with $%
\protect\alpha=0.265005$ and $\protect\beta=0.800925$. ``dS'' stands for de
Sitter universe.}
\label{fig8}
\end{figure}

We observe the trajectories of the hybrid scale factor obtained in our model
in Figs.\ref{fig7}-\ref{fig8}, in which the values of $\alpha$ and $\beta$
were chosen in accordance to the previous section results. They follow a
pattern similar to that of power law cosmology. The trajectories in $s-r$
plane in the power law cosmology (see Ref.\cite{Sarita}) start from SCDM
(standard cold dark matter ($\Lambda=0$)) point and follows up the $\Lambda $%
CDM model while this hybrid model is deviated from the SCDM and follows up
the $\Lambda $CDM model. Similarly, the trajectories in the $q-r$ plane can
be seen and compared with the power law cosmology.


\section{Discussion}\label{sec:d}

The longstanding cosmological constant problem remains as one of the
greatest observational issues of Physics. An interesting form to alleviate it is by considering the cosmological \textquotedblleft constant\textquotedblright\
as time varying (for instance, Ref.\cite{basilakos/2009}). Besides changing
the gravitational aspects of gravity, as deeply mentioned in Section \ref%
{sec:int}, one can also infer the cosmic acceleration by assuming the
Universe is filled with some fluid that follows the Chaplygin gas
equation of state \cite{bento/2002}-\cite{bento/2003b}. This is also an
alternative to describe cosmic acceleration with no cosmological constant,
and in fact it can provide an unified alternative to dark energy and dark
matter.

Here we have addressed the cosmic acceleration through an
extended theory of gravity named $f(R,T)$ gravity that allows the inception of material terms dependent on $T$ in the gravitational action. As a novelty in the literature, we considered an exponential dependence for the trace of the energy-momentum tensor.

We have used here the Hubble parameter data obtained by different age techniques applied to evolving galaxies and constrained the model parameters $\alpha $ and $\beta $ in our model respectively as $\alpha =0.265005$ and $\beta =0.800925$, with minimum $\chi ^{2}=31.6247$ based on Bayesian statistics. Moreover, we have obtained the likelihood contours at $1\sigma $, $2\sigma $ and $3\sigma $ confidence levels using the OHD as shown in the $\alpha -\beta $ plane of Figure \ref{fig:Contour-EXP}. We have compared our model with the $\Lambda $CDM model. We found that our model is fitting well the OHD with the constrained values of $\alpha$ and $\beta$ and does not deviate significantly from $\Lambda$CDM (see Figure \ref{fig:hzerror}), though for higher values of redshift the present model clearly presents a better adjust with observations.

The statefinder diagnostic, which was here obtained for the $R+e^T$ cosmology,
is generally used to compare various dark energy models and their deviations
from $\Lambda$CDM and SCDM models. We have shown the time and redshift
evolutions of various cosmological parameters. We have demonstrated the
phase transition redshift through graphical representation (Fig.\ref{fig1}).
All of these features elevate the $R+e^T$ cosmology to the level of an
important alternative to describe the cosmic acceleration with no need of a
cosmological constant.

Finally, by power expanding $e^T$ in the model action, one obtains $R+T+1+\Theta(T^2)$. In the first order approximation and properly inserting the right units, the third term in the expansion reads as a cosmological constant, which naturally appears in the model. Since the $f(R,T)=R+T$ cosmology has been discarded from observational data \cite{velten/2017}, this expansion appears as a good alternative to revive this scenario.

\section*{Acknowledgements}

PHRSM would like to thank CAPES for financial support. PKS acknowledges CSIR, New Delhi, India for financial support to carry out the Research project
[No.03(1454)/19/EMR-II Dt.02/08/2019].


\begin{thebibliography}{99}

\bibitem{hubble/1929} E. Hubble, Proc. Nat. Acad. Sci. U.S.A. \textbf{15}
(1929) 168-173.

\bibitem{riess/1998} A.G. Riess et al., Astron. J. \textbf{116} (1998)
1009-1038.

\bibitem{perlmutter/1999} S. Perlmutter et al., Astrophys. J. \textbf{517}
(1999) 565-586.

\bibitem{planck_collaboration/2017} Planck Collaboration: N. Aghanim et al.,
Astron. Astrophys. \textbf{607} (2017) A95.

\bibitem{weinberg/1989} S. Weinberg, Rev. Mod. Phys. \textbf{61} (1989) 1.

\bibitem{peebles/2003} P.J. Peebles, B. Ratra, Rev. Mod. Phys. \textbf{75}
(2003) 559.

\bibitem{padmanabhan/2003} T. Padmanabhan, Phys. Rep. \textbf{380} (2003)
235-320.

\bibitem{bednyakov/2016} V.A. Bednyakov, Is it possible to discover a dark
matter particle with an accelerator?, Phys. Part. Nucl. \textbf{47} (2016)
711-774. https://doi.org/10.1134/S1063779616050026

\bibitem{mambrini/2007} Y. Mambrini, E. Nezri, Eur. Phys. J. C \textbf{50}
(2007) 949-968.

\bibitem{einstein/1915} A. Einstein, Sitzungsber. Preuss. Akad. Wiss.
Berlin(1915) 844-847.

\bibitem{de_felice/2010} A. De Felice, S. Tsujikawa, Liv. Rev. Rel. \textbf{%
13} (2010) 3.

\bibitem{capozziello/2011} S. Capozziello, M. de Laurentis, Phys. Rep. 
\textbf{509} (2011) 167-321.

\bibitem{nojiri/2006} S. Nojiri et al., Phys. Rev. D \textbf{74} (2006)
046004.

\bibitem{shirasaki/2016} Y. Shirasaki et al., Publ. Astron. Soc. Jap. 
\textbf{68} (2016) 23.

\bibitem{capozziello/2013} S. Capozziello et al., J. Cosm. Astrop. Phys. 
\textbf{2013} (2013) 024.

\bibitem{rodrigues/2014} D. C. Rodrigues et al., Month. Not. Roy. Astron.
Soc. \textbf{445} (2014) 3823-3838.

\bibitem{chiba/2003} T. Chiba, Phys. Lett. B \textbf{575} (2003) 1-3.

\bibitem{nojiri/2008} S. Nojiri, S.D. Odintsov, Phys. Lett. B \textbf{659}
(2008) 821-826.

\bibitem{harko/2011} T. Harko et al., Phys. Rev. D \textbf{84} (2011) 024020.

\bibitem{odintsov/2017} S.D. Odintsov et al., Eur. Phys. J. C \textbf{77}
(2017) 862.

\bibitem{pettorino/2005} V. Pettorino et al., J. Cosm. Astrop. Phys. \textbf{%
2005} (2005) 014.

\bibitem{harko/2010} T. Harko, F.S.N. Lobo, Eur. Phys. J. C \textbf{70}
(2010) 373.

\bibitem{oliveira/2015} A. M. Oliveira et al., Phys. Rev. D \textbf{92}
(2015) 044020.

\bibitem{harko2/2011} T. Harko et al., Mod. Phys. Lett. A \textbf{26} (2011)
1467.

\bibitem{mignani/1997} R. Mignani et al., Gen. Rel. Grav. \textbf{29} (1997)
1049-1073.

\bibitem{harko/2015} T. Harko et al., Eur. Phys. J. C \textbf{75} (2015) 386.

\bibitem{harko/2014} T. Harko, Phys. Rev. D \textbf{90} (2014) 044067.

\bibitem{steigman/2009} G. Steigman et al., An accelerating cosmology
without dark energy, J. Cosm. Astrop. Phys. \textbf{2009} (2009) 033.
https://doi.org/10.1088/1475-7516/2009/06/033

\bibitem{lima/2010} J.A.S. Lima et al., J. Cosm. Astrop. Phys. \textbf{1011}
(2010) 027.

\bibitem{lima/2012} J.A.S. Lima et al., Phys. Rev. D \textbf{86} (2012)
103534.

\bibitem{friedmann/1922} A. Friedmann, Phys. A \textbf{10} (1922) 377\^{a}%
\euro ``386.

\bibitem{friedmann/1924} A. Friedmann, Zeits. Phys. A \textbf{21} (1924) 326%
\^{a}\euro ``332.

\bibitem{lemaitre/1927} G. Lema\^itre, Annal. Soc. Scient. Brux. \textbf{A47}
(1927) 49-59.

\bibitem{robertson/1935} H.P. Robertson, Astrophys. J. \textbf{82} (1935)
284.

\bibitem{walker/1937} A.G. Walker, Proc. Lon. Math. Soc. \textbf{42} (1937)
90-127.

\bibitem{Hz} G. S. Sharov, V. O. Vasiliev, Mathematical Modelling and
Geometry \textbf{6(1)} (2018) 1-20.

\bibitem{planck_collaboration/2015} Planck Collaboration, Astron. \&
Astrophys. \textbf{594} (2016) A13.

\bibitem{Sahni/2003} V. Sahni et al., JETP Lett. \textbf{77} (2003) 201-206.

\bibitem{Alam/2003} U. Alam et al., Mon. Not. R. Astron. Soc. \textbf{344}
(2003) 10571074.

\bibitem{Sarita} Sarita Rani et al., J. Cosmol. Astropart. Phys. \textbf{2015%
} (2015) 031.

\bibitem{Capozziello14} S. Capozziello, O. Farooq, O. Luongo, B. Ratra,
Phys. Rev. D, \textbf{90} (2014) 044016.

\bibitem{Capozziello15} S. Capozziello, O. Luongo, E. N. Saridakis B., Phys.
Rev. D, \textbf{91} (2015) 124037.

\bibitem{Farooq17} O. Farooq, F. Madiyar, S. Crandall, B. Ratra, Astrophys.
J., \textbf{835} (2017) 26.

\bibitem{giostri/2012} R. Giostri et al., J. Cosm. Astrop. Phys. \textbf{03}
(2012) 027.

\bibitem{basilakos/2009} S. Basilakos et al., Phys. Rev. D, \textbf{80}
(2009) 083511.

\bibitem{bento/2002} M.C. Bento et al., Phys. Rev. D, \textbf{66} (2002)
043507.

\bibitem{gorini/2003} V. Gorini et al., Phys. Rev. D, \textbf{67} (2003)
063509.

\bibitem{debnath/2004} U. Debnath et al., Class. Quant. Grav. \textbf{21}
(2004) 5609-5618.

\bibitem{bento/2003} M.C. Bento et al., Phys. Rev. D, \textbf{67} (2003)
063003.

\bibitem{bento/2003b} M.C. Bento et al., Phys. Lett. B \textbf{575} (2003)
172-180.

\bibitem{velten/2017} H. Velten and T.R.P. Caram\^es, Phys. Rev. D, {\bf 95} (2017) 123536.

\end{thebibliography}
\end{document}